\documentclass[twocolumn]{aastex631}

\newcommand{\nubhlight}{$\nu$\texttt{bhlight}~}
\newcommand{\HARM}{\texttt{HARM}~}
\newcommand{\bhlight}{\texttt{bhlight}~}

\graphicspath{{./}{figures/}}

\usepackage{amsmath}
\usepackage{array}

\begin{document}

\title{The Dependence of Gamma-Ray Burst Jet Collimation on Black Hole Spin}

\author[0000-0002-3221-9395]{Valeria U. Hurtado}
\affiliation{University of Washington. 3910 15th Ave NE. Seattle, WA 98195. USA.}
\affiliation{Fisk University. 1000 17th Ave N. Nashville, TN 37208. USA.}
\affiliation{Computational Division, Los Alamos National Laboratory, Los Alamos, NM 87545, USA}

\author[0000-0003-1707-7998]{Nicole M. Lloyd-Ronning}
\affiliation{Computational Division, Los Alamos National Laboratory, Los Alamos, NM 87545, USA}
\affiliation{Dept of Math, Science and Engineering, University of New Mexico, Los Alamos, NM 87545, USA}
\affiliation{Center for Theoretical Astrophysics, Los Alamos National Laboratory, Los Alamos, NM 87545, USA}

\author[0000-0001-6432-7860]{Jonah M.\ Miller}
\affiliation{Computational Division, Los Alamos National Laboratory, Los Alamos, NM 87545, USA}
\affiliation{Center for Theoretical Astrophysics, Los Alamos National Laboratory, Los Alamos, NM 87545, USA}



\begin{abstract}

Gamma-Ray Bursts are the most luminous events in the Universe, and are excellent laboratories to study extreme physical phenomena in the cosmos. Despite a long trajectory of progress in understanding these highly energetic events, there are still many observed features that are yet to be fully explained. Observations of the jet opening angle of long gamma-ray bursts (LGRBs) suggest that LGRB jets are narrower for those GRBs at higher redshift.  This phenomenon has been explained in the context of collimation by the stellar envelope, with denser (lower metallicity) stars at higher redshifts able to collimate the jet more effectively. However, until now, the dependence of jet opening angle on the properties of the central engine has not been explored.  We investigate the effect of black hole spin on the jet collimation angle for a magnetically launched jet, using the General Relativistic Radiation Magnetohydrodynamical (GRRMHD) code \nubhlight. We present 3D results for a range of spin values. The simulations show that higher spinning black holes tend to create narrower jets. If indeed LGRB progenitors in the early universe are able to produce black hole central engines with higher spin, this could account for at least some of the observed jet opening angle-redshift correlation.

\end{abstract}



\section{Introduction} \label{sec:intro}

Gamma-Ray Bursts (GRBs) are the most powerful events in the Universe, releasing energies in the order of $10^{52}$ erg in a typical event \cite[for reviews see, e.g.][]{2004RvMP...76.1143P,ZM04, Mesz06, 2009ARA&A..47..567G, DAvanz15,Lev16}. They are bursts of gamma-rays that can last from less than a second to several thousands of seconds depending on the progenitor. For GRBs lasting 2 seconds or less, they are thought to come from the merger of two neutron stars; see, e.g. \cite{2007NJPh....9...17L,2014ARA&A..52...43B}. If the gamma-ray emission lasts from a few seconds up to hundreds of seconds, they are thought to come from a hydrogen deficient star that collapses into a black hole (BH)\footnote{There also exist viable GRB models with a magnetar central engine \citep{Usov92,DT92,Thomp94,ZM01} but here we consider only a black hole central engine.}, often termed a collapsar \citep{2006ARA&A..44..507W,2012ApJ...752...32W}. Our motivation for this work comes from observational results of the latter, so-called long gamma-ray bursts (LGRBs).  However, our results apply to any GRB black hole-accretion disk central engine.

A LGRB begins with a star whose core collapses into a BH. This, along with the remaining gas available from the collapse of the star that is gravitationally bound to the newly formed BH and able to form an accretion disk, are necessary elements to give rise to the LGRB central engine. \citep{2001ApJ...550..410M,2006ApJ...637..914W}.  As general relativistic frame-dragging causes more rapid rotation near the black hole, magnetic fields present in the disk will be wound up along the spin axis of the black hole. As a result, a powerful Poynting flux is generated along this axis and a highly relativistic jet is launched through the so-called Blandford-Zjanek (BZ mechanism)  \citep{1977MNRAS.179..433B, MT82}. This is an efficient way to form and power a jet, assuming there exists enough angular momentum and magnetic flux in the system.  
The BZ mechanism does not depend on the mass accretion rate (at least not until the disk reaches a MAD state), but instead, again, on the rotation of the BH and the presence of magnetic flux. For additional detailed discussion of BZ jets in the context of GRBs, see \cite{LWB00,LRR02,tm12,Lei17, LR19b}.

Gamma-Ray Bursts still pose many mysteries - from the nature of their progenitors to the details of the particle acceleration and radiation mechanisms in their relativistic jets - and multiple studies continue to uncover novel behaviors in these energetic events. In one such example, \cite{LR19} found a statistically significant anti-correlation between jet opening angle and redshift (with narrower jets at higher redshifts) in a large sample of LGRBs \cite[we note the presence of this potential correlation first emerged in the analysis of \cite{Frail01}, and has also been suggested, although not to the same statistical significance, in][]{LRFRR02, Yon05, Lu12, Las14, Las18}. Conservatively accounting for the presence of any potential selection effects or observational biases that may produce the correlation (using the well-tested and vetted non-parametric Lynden-Bell \citep{LB71} and Efron-Petrosian \citep{EP92} methods), they found that jet beaming angle goes to $\theta_{j} \propto (1+z)^{-0.75 \pm 0.25}$.

In a following study, \cite{Lloyd_Ronning_2020} provided potential reasons as to why we may see a difference in jet beaming angle over cosmic redshift. They specifically looked into how some of the properties of the progenitor star might affect the jet opening angle. While looking at the luminosity requirements to launch a successful jet, \cite{Lloyd_Ronning_2020} 
found that the anti-correlation between jet opening angle and redshift can be explained by an evolution of the progenitor envelope over cosmic time. At highter redshifts, lower metallicity, denser progenitors can more effectively collimate the jet as it traverses the stellar envelope and cocoon region. This is quantitatively consistent with evolving IMF models reported in the literature \cite[e.g.][]{Dave08}. 


However, the role of the central engine in producing this correlation has yet to be explored.  In particular, the BH-accretion disk dynamics are not yet well understood in relation to the jet geometry. There have been many works, beginning with \citet{McKinney2004}, exploring the relationship between jet power (with an eye to AGN luminosity) and black hole and disk conditions \citep{tm12,tmn12}. However, less attention has been paid to opening angle. How this property depends on redshift is an additional aspect necessary to understanding the opening angle-redshift anti-correlation, and LGRBs in general. For instance, a progenitor at higher redshift may potentially lose less angular momentum over the lifetime of the star, due to its lower metallicity compared to progenitors at lower redshifts (assuming angular momentum loss comes from radiation-driven stellar winds, which are stronger at higher metallicity). This could eventually lead to a remnant with higher angular momentum as well. Therefore, a progenitor that collapses at higher redshift with higher angular momentum (compared to one at a lower redshift) may produce a more highly spinning BH with a narrower jet.  

We study this possibility using a General Relativistic Radiation Magnetohydrodynamical (GRRMHD) code, \nubhlight. With \nubhlight we simulate a BH with an accretion disk and a jet in 3 dimensions and let it evolve in time. Our goal is to consider the degree of collimation of a Blandford-Znajek jet \citep{BZ77,MT82} as a function of the spin of the black hole. Recently \citet{2022MNRAS.511.3795N} performed a detailed examination of jet power and geometry for magnetically arrested disks; this differs from our work here - we study non-arrested (SANE) disks.

\begin{figure}[t]
    \centerline{ 
  \includegraphics[width=4.0in]{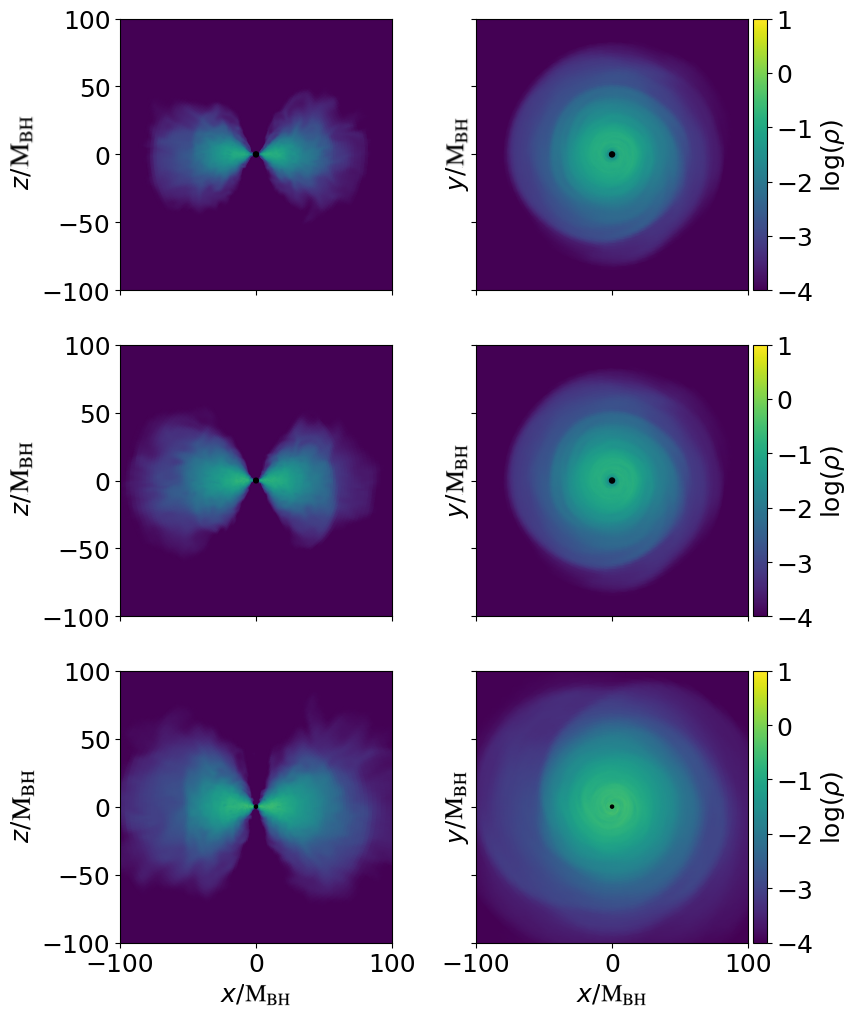}
    } 
\caption{{\label{fig:3d_0p50p99}} 3D simulations showing a plot of density for 3 simulation runs, all at time 5500 $\rm M_{\rm BH}$ out of 10000 $\rm M_{\text{BH}}$. On the top is the simulation corresponding to a BH of spin 0.5, in the middle is a BH of spin 0.7, and on the bottom a BH of spin 0.99. All simulations show a frontal view of the black hole - accretion disk - jet (bhadj) system. The blue and purple hues correspond to the least dense material (jet) in the system, while green and yellow hues to the densest (accretion disk).}
\end{figure}

This letter is organized as follows: In \S 2, we describe the GRMHD code \nubhlight, and the set up we used in our simulations.  In \S 3, we present the results of our 3D simulations for different spin values of of the black hole. We show that, using various ways to quantify the jet opening angle, {\em more highly spinning black holes tend to produce narrower jets}. In \S 4, we summarize our findings and discuss both the caveats and the implications of our results in the context of our current understanding of long gamma-ray bursts.

\begin{figure}[t]
  \centerline{ 
  \includegraphics[width=2.5in]{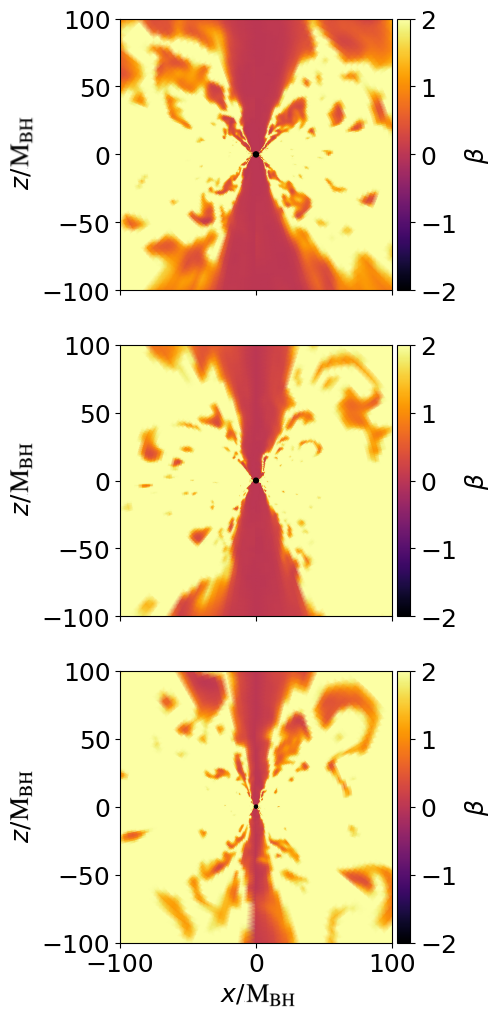}}
\caption{{\label{fig:3d_beta}} 3D simulations of the BH - accretion disk - jet system, all at time 5500 $\rm M_{\rm BH}$ out of 10000 $\rm M_{\rm BH}$. On the top is a BH with spin 0.5 (lower spin fraction), in the middle is a BH of spin 0.7, and on the bottom a BH of spin 0.99 (highest spin fraction). $\beta >$ 1 indicates material dominated by gas pressure (accretion disk - in yellow/orange) and $\beta <$ 1 indicated material dominated by magnetic field energy (jet - pink/purple). }
\end{figure}


\section{Numerical Tools} \label{sec:style}

For our simulations we use \nubhlight (read as ``nublight"). \nubhlight is a General Relativistic Radiation Magnetohydrodynamics (GRRMHD) code, built on top of the \bhlight \citep{ryan2015bhlight}, \texttt{grmonty} \citep{Dolence_2009}, and \HARM \citep{Gammie_2003} codes. In this work, we disable the radiation transport and solve only the equations of general relativistic ideal magnetohydrodynamics (GRMHD). These include conservation of Baryon number:
\begin{equation}\label{barnum}
\partial_t (\sqrt{-g}\rho_o u^t) \; + \; \partial_i (\sqrt{-g}\rho_o u^i) \; = \; 0,
\end{equation}
conservation of energy and momentum,
\begin{equation}\label{enmomcons}
\begin{split}
    \partial_t [\sqrt{-g} & \, ( \, T_\nu^t + \rho_o u^t \delta_\nu^t \, ) \, ] \; + \; \partial_i [\sqrt{-g} \, ( \,  T_\nu^t + \rho_o u^i \delta_\nu^i \, ) \, ] = \\
    & \sqrt{-g} \, T_\lambda^\kappa \, \Gamma_{\nu \kappa}^\lambda \; \; \forall \; \; \nu = 0,1...3,
\end{split}
\end{equation}
and conservation of magnetic flux,
\begin{equation}\label{evomagfield}
    \partial_t \, (\sqrt{-g} \, B^i \,) \, - \, \partial_j \, [\sqrt{-g} \,  (b^j u^i - b^i u^j \, ) \, ] \, = \, 0 ,
\end{equation}
given the energy-momentum tensor defined as 
\begin{equation}\label{energytens}
    \begin{split}
    T_\nu^\mu \, = \, (\rho_o + P + b^2 \, ) \, u^\mu u_\nu \,  + \, \partial_\nu^\mu \, ( \, P \, + 1/2 \, b^2 \, ) \, - \, b^\mu \, b_\nu,
    \end{split}
\end{equation}
for metric $g_{\mu \nu}$, rest energy $\rho$, fluid four velocity $u^{\mu}$, internal energy density $u$, pressure $P$, and Christoffel symbol $\Gamma^\alpha _{\beta \gamma}$. In the equations above, we also make use of the magnetic field components $B^i = \mbox{*} F^{it}$ of the Maxwell tensor $\mbox{*} F_{\mu \nu} = b^\mu u^\nu \, - \, b^\nu v^\mu$, and the magnetic field four-vector $b^\mu$. Here and throughout the text we use units of $G=c=1$ and standard Einstein index notation, where Greek indices range from 0 to 3 for spacetime indices and roman indices range from 1 to 3 for spatial indexes only. We also subtract off Equation \eqref{barnum} from the zeroth component of Equation \eqref{enmomcons} to remove the rest energy from the energy conservation law.

The fluid equations \eqref{enmomcons} and \eqref{energytens} must be closed by an \textit{equation of state}. In this work we use the standard ideal gas equation of state 
\begin{equation}
    \label{eq:ideal:gas}
    P = (\Gamma - 1) u
\end{equation}
with $\Gamma=5/3$ corresponding to an ionized gas. All simulations performed assume a stationary Kerr black hole background \citep{1963PhRvL..11..237K}. The \nubhlight GRMHD treatment includes the industry standard treatments for codes of the \HARM family, including the radially logarithmic quasi-spherical grid in horizon penetrating coordinates, as first described in \citet{McKinney2004}, the WENO reconstruction first described in \citet{Tchekhovskoy2007}, the primitive variable recovery scheme described in \citet{MignoneMckinney2007}, and the drift-frame artificial atmosphere treatment described in \citet{Ressler2015ebhlight}. There is a long history of GRMHD simulations in the community. For a recent discussion of issues and methods, see \citet{EHTComparison2019}.

We run six identical simulations where the only difference between them lies on the black hole spin $\rm a_{\rm BH}$ values used $\rm a_{\rm BH}$ = 0.4, 0.5, 0.7, 0.85, 0.95, 0.99 (the maximum corresponds to $\rm a_{\rm BH}$ = 1). We begin each simulation with a torus in hydrostatic equilibrium \citep{1976ApJ...207..962F} with an inner radius of $\rm r_{\rm in} =  6 G M_{\rm BH}/c^2$ and a radius of maximum pressure of $\rm r_{\rm max,P} = 12 G M_{\rm BH}/c^2$. 

Because the ideal gas equation of state (EOS) has no fundamental energy scale, the disk-accretion problem becomes scale free, and our problem setup is valid for a large range of black hole masses and for accretion rates roughly corresponding to a maximum rate of a $\sim 1$ in dimensionless units (one $\rm M_{\rm BH}/(GM_{\rm BH}/c^3)$). We thread our inital torus with a single poloidal magnetic field loop such that the ratio of gas to magnetic pressure
\begin{equation}\label{betap}
    \beta = \frac{P_{\rm gas}}{P_B}
\end{equation}
is 100 at the radius of maximum pressure.
Where here 
\begin{equation}
    P_B = \frac{B^2}{2}.
\end{equation}
 We choose this $\beta$ so that the initial value is sufficiently weak that the magnetic pressure does not contribute to the support of the equilibrium torus. This weak value also allows the magnetic field to grow self-consistently from turbulence and minimize artifacts of the initial data.
We use a resolution of  $\rm N_r$ x $\rm N_\theta$ x $\rm N_\phi$ = 256 x 192 x 128.

Molecular viscosity in the disk is negligible. As the disk evolves, the magnetorotational instability (MRI) \citep{VelikovMRI,BalbusHawley91} enhances the magnetic field strength and drives turbulences, which in turn provides turbulent viscosity, enabling angular momentum transport and accretion \citep{ShakuraSunyaev1973}. This same magnetic field drives a jet through the BZ mechanism \citep{1977MNRAS.179..433B}.  We run our simulations for a duration of $\rm t =10^4 G M_{\rm BH}/c^3$, long enough for the accretion flow to achieve a quasi-stationary flow. 

 For this flow structure to be trustworthy, the fastest growing mode of the MRI must be captured. We use the MRI quality factor defined in \citet{Miller_2019} and find similar structure to the quality factor reported there, with $Q^{(\theta)}_{\texttt{MRI}}$ decaying over time. In these models, we use the equation of state of an ideal gas. \citet{Miller_2019} uses a nuclear equation of state, which has a high enthalpy, making the wavelength of the MRI shorter. This increases the quality factor so that $Q^{(\theta)}_{\texttt{MRI}} > 10$ for all times. The community has performed a battery of convergence tests over the years \citep{Sano2004,Shiokawa2011}, culminating in a large code-comparison effort for the Event Horizon collaboration \citep{EHTComparison2019}, and the industry standard for a ``resolved'' simulation is a $Q^{(\theta)}_{\texttt{MRI}} > 10$. It is known that the effective resolution in \HARM \citep{Gammie_2003} codes is enhanced by the focusing of the curvilinear grid. Moreover a WENO \cite{Tchekhovskoy2007} reconstruction (which we apply) adds an additional factor of approximately 5 in effective resolution, thanks to reduced numerical viscosity.

\begin{figure*}[t]
  \centerline{ 
  \includegraphics[width=3.2in]{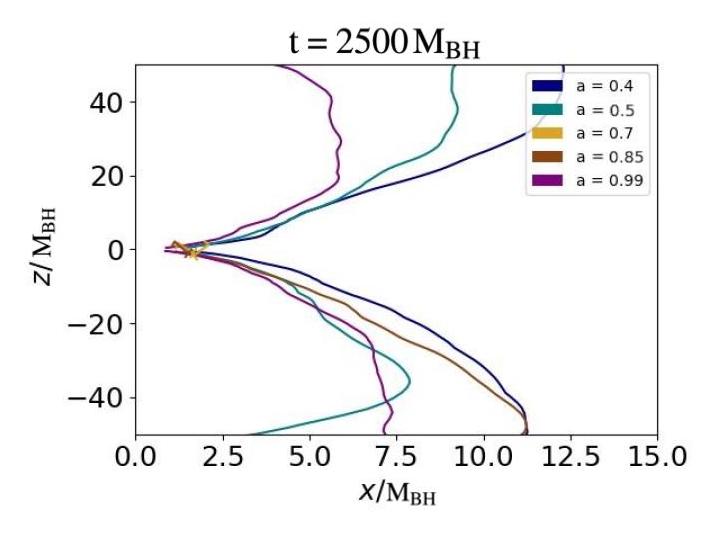}
  \includegraphics[width=3.2in]{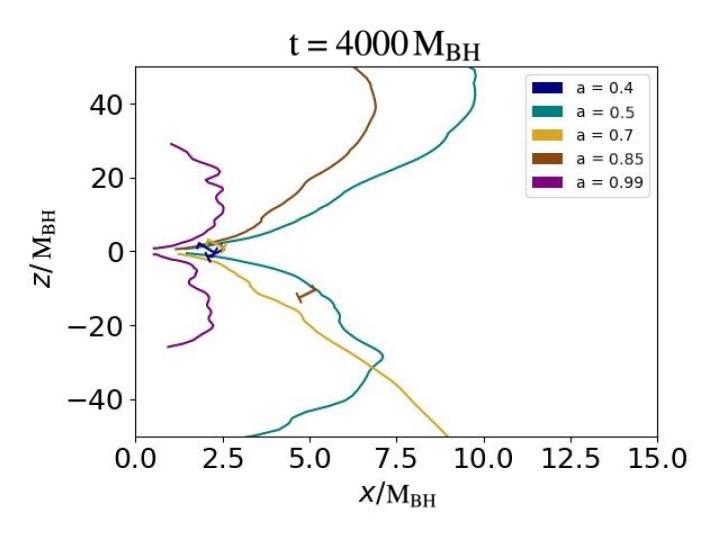}} 
    \centerline{ 
  \includegraphics[width=3.2in]{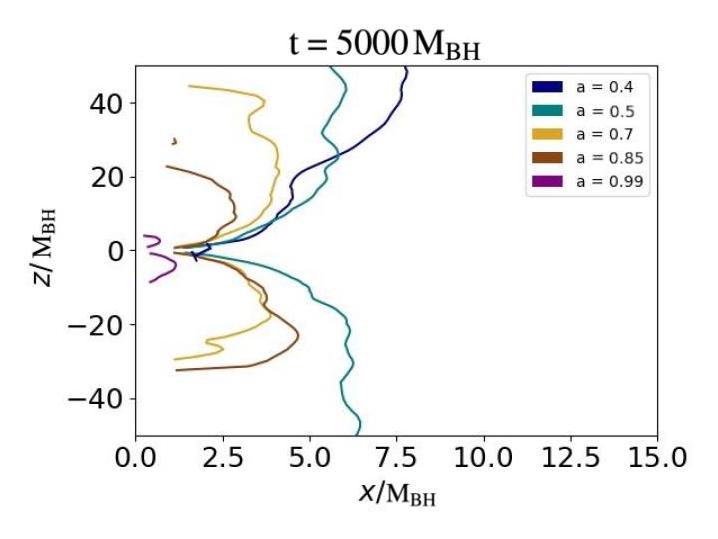} 
  \includegraphics[width=3.2in]{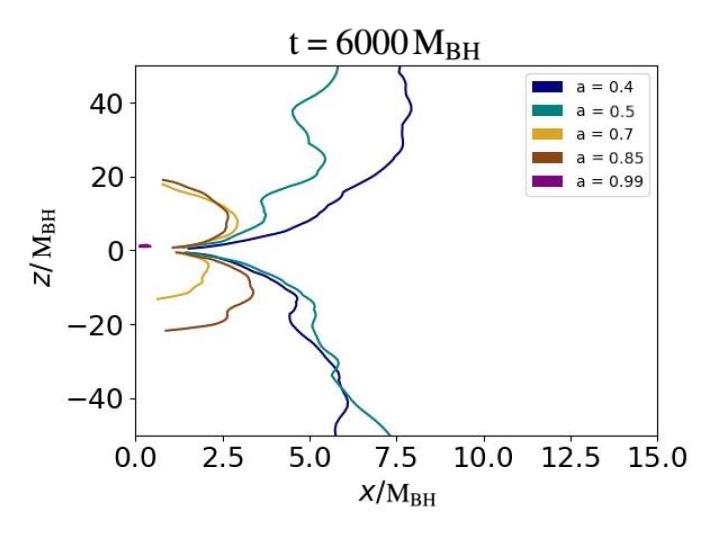}}
\caption{{\label{fig:3dsigma}} Plots of 3D simulation sigma ($\sigma$) analysis in different times of the simulation. Each plot is a snapshot in time of $\sigma$ for the case of 5 different BH spins. The navy curve corresponds to the BH of spin 0.4, the teal to a BH of spin 0.5, the yellow curve  to the BH of spin 0.7, the brown to a BH of spin 0.85, and the purple line to the BH of spin 0.99. The top left plot is at the earliest time, and the bottom right is at a later time in the simulation. This plot indicates the trend that a more highly spinning BH leads to a narrower jet opening angle and vice versa.}
\end{figure*}

\section{Results} \label{sec:floats}

Figure \ref{fig:3d_0p50p99} shows the density profiles of the black hole--accretion disk--jet system for three different 3D simulation runs. On the top panel is the simulation corresponding to a BH of spin 0.5, the middle to a BH of spin 0.7, and on the bottom a BH of spin 0.99. The left column shows poloidal slices, while the right shows equatorial.  We see significant magnetically dominated outflows at the poles which we discern as the jet (defined in more detail below).

While jet structure is more challenging to discern on plots of density, the accretion disk on both of these simulations appears to differ. For the BH with 0.99 spin (bottom of Figure \ref{fig:3d_0p50p99}), the accretion disk appears to be much larger in size (about twice the size) than the simulation with 0.5 spin, for instance. The structure of the accretion disk can highly affect jet opening angle, as previously mentioned. A larger, puffier accretion disk may provide further collimation of the jet, since the jet will have to spend more time traveling through the accretion disk material. This could lead to a narrower jet opening angle.   We note that, importantly, the initial conditions of the disk depend on the black hole spin \citep{1976ApJ...207..962F}.  For example, for a Keplerian velocity profile in the disk, higher initial black hole spins will force a slightly thicker accretion disk in the Fishbone-Moncreif setup we employ (see, e.g., their Figure 2). We discuss this further in \S 4.

To obtain a better view of the jet structure, we calculated $\beta$ for all simulations in Figure \ref{fig:3d_beta}, where $\beta >$ 1 indicates material dominated by gas pressure (shown in yellow/orange colors representing the gas in the accretion disk) and $\beta <$ 1 indicates material dominated by magnetic field energy (shown in pink/purple hues representing the jet). Thus, the accretion disk would be the area dominated by gas pressure, and the jet would be the material dominated by magnetic field pressure. The top of figure \ref{fig:3d_beta} shows the BH of 0.5 spin, the middle a BH of spin fraction of 0.7, and the bottom panel shows the BH of spin 0.99. From a first glance, one can see that the jet for the BH of spin 0.99 seems to be much narrower than for the BH of spin 0.5. Qualitatively, the higher-spin run shows a more collimated jet than the lower spin runs. 

\begin{table}
    \centering
        \caption{Table of the average sigma weighted angle ($\overline{\theta}_\sigma$) values as a function of spin. The angle $\theta$ is measured from the equator to the `edge' of the jet. Therefore, a greater value of $\overline{\theta}_\sigma$ signifies a narrower jet opening angle The averages and their standard deviations have been computed across ten different times in the simulation ranging from 4000 $\rm M_{BH}$ to 6000 $\rm M_{BH}$. The values indicate, on average, a wider jet opening angle for more slowly spinning BHs, and a narrower jet opening angle for the more rapidly spinning BHs, corroborating what is seen by eye in Figure 3.}
     \begin{tabular}{cc}
    \hline 
        
        \hline
        \hline
         $\rm a_{BH}$&   $\rm \overline{\theta}_\sigma$ [deg]\\ [1ex]  \hline 
           0.4&   73.4 $\pm$ 0.1\\ 
         0.5&   72.8 $\pm$ 0.2\\ 
         0.7&  74.9 $\pm$ 0.3\\ 
         0.85&   76.5 $\pm$ 0.7\\ 
         0.95&   76.6 $\pm$ 0.6\\ 
         0.99&   75.9 $\pm$ 0.4 \\ 
\hline
\hline

    \end{tabular}

    \label{tab:angles}
\end{table}


We also examine the magnetization parameter sigma
\begin{equation}
    \sigma = B^2 / (2 \rho c^{2}),
\end{equation}
which is the ratio of the magnetic energy density to rest mass energy density. We make this choice for a number of reasons: not only does $\sigma$ provide a good estimate of where magnetic energy dominates in our (magnetically-launched) outflows, but it is better defined in the jet region than $\beta$. The quantity $\beta$ requires a temperature estimate of the gas, but the outflow is expected to be collisionless in the region around the spin axis of the black hole and consequently temperature is not well defined. Therefore we suggest $\sigma$ is a better measure and representation of the jet structure and edge. 
We plot isocontours of $\sigma = 1$ in a poloidal slice for our runs at four different times in Figure \ref{fig:3dsigma}, at $\rm t=2500 \, M_{BH}$, $\rm 4000 \, M_{BH}$, $\rm 5000 \, M_{BH}$ and $\rm 6000 \, M_{BH}$. Each plot is a snapshot in time of $\sigma$ for all 3 of the BH spins, plus an additional two runs of BH spin $\rm a_{BH} = 0.4, 0.85$. The navy curve corresponds to a BH spin of 0.4, the light blue curve to the BH of spin 0.5, the yellow curve corresponds to the BH of spin 0.7, the brown curve to a spin of 0.85, and the purple line to the BH of spin 0.99. The snapshots correspond to sequential times in the simulation. In other words, the top-right plot is at an earlier time in the simulation, top-left at a slightly later time, and so on. 
As with Figure \ref{fig:3d_beta}, more highly spinning black holes have more collimated jets. 

To fully quantify this trend, we calculate the average disk angle $\theta$ weighted by $\sigma$,
\begin{equation}
    \overline{\theta}^2_\sigma = \frac{\int_\Omega\sqrt{-g} \; \theta^2 \, \sigma \; dx^1 \, dx^2 \, dx^3}{\int_\Omega \sqrt{-g} \; \sigma \; dx^1 \, dx^2 \, dx^3},
    \label{thtsig}
\end{equation}
Where $\sqrt{-g}$ is the square root of the determinant of the metric, $\theta$ is the angle measured from the equator of the simulation, and $dx^1$, $dx^2$, $dx^3$ the measure. This integral is performed over the whole simulation domain. This acts as a proxy for jet opening angle: a larger $\overline{\theta}_\sigma$ means a narrower jet beaming angle. The result of this calculation is shown in Table \ref{tab:angles}. In this table we have also included additional simulation runs of $\rm a_{BH} =$ 0.4, 0.85, and 0.95 for completeness. 
While the difference between jets may seem small, it follows the general trend that has been seen throughout all 3D simulations alike: there is a difference in jet collimation as a function of spin. More specifically, there seems to be a trend where more highly spinning BHs lead to narrower jet opening angles. 


\section{Discussion \& Conclusions}
Using the GRMHD code \nubhlight we have investigated the dependence of the jet opening angle on the black hole spin. We find that, on average, more rapidly spinning black holes lead to narrower jets.  We can see this by eye from the snapshots of our 3D numerical solutions (Figures 1 and 2), where the effect appears fairly pronounced.  It is also borne out by our more detailed analysis of the contours of both plasma $\beta$ and $\sigma$ (Figure 3 and Table \ref{tab:angles}),  although in the latter case of the $\sigma$-weighted average angle, the change in angle with spin is less pronounced and appears to saturate at high values of the spin. We see this pattern track across simulations and spins.  We emphasize again the scale-free nature of our simulations; as such, these results should apply to a wide range of black hole-disk systems including those relevant to short GRBs and even supermassive black hole-disk systems. 

As noted in \S 3, it is important to consider that our initial disk set up based on \cite{1976ApJ...207..962F} depends on black hole spin in a nuanced way, with slightly thicker initial disks for more highly spinning black holes.  This may contribute to the ``narrowness'' of the jet at later times in our simulation.   Indeed other studies have shown how the initial conditions may influence the jet over the simulation time, including how the torus size (which is spin dependent) can influence the jet opening angle \citep{Lisk18, Chris19, Roh23}. That said, the set-up - a torus in hydrostatic equilibrium - is a solution to the relativistic Euler equations and therefore provides a physical framework for this result in some sense.  That is, an initially thicker disk may lead to a narrower jet, but this is physically motivated and a reflection of the balance between magnetic, gravitational and gas pressures (not only initially but as the system evolves).  Additionally, consider that a more rapidly rotating black hole will have a horizon radius (and innermost stable circular orbit) that is closer to the black hole.  This allows magnetic flux to be brought into a narrower region near the black hole (i.e. a smaller polar angle) and may also contribute to the qualitative physical explanation for more rapidly spinning black holes producing narrower jets.

Nonetheless, these results are broadly consistent with the analysis for magnetically arrested disks in \citet{2022MNRAS.511.3795N}, which help further connect the GRB central engine and jet properties, and may help us better understand the underlying progenitor(s) of GRBs.   Ultimately, the collimation will depend on the equilibrium between the external thermal pressure and the magnetic pressure of the Poynting flux of the jet as shown in \cite[][see also \cite{Lyub09,Kom09}]{Pais2023}. 

Another confounding factor is the uncertainty in the initial magnetic field strength. Our simulations are constructed so that the magnetic pressure at initial times is not a significant contributor to the pressure support of the disk. As the disk evolves, the MRI drives growth in the field strength until it is roughly in equipartition. A stronger initial magnetic field will both contribute to the initial disk pressure and shrink the amount of time required for the MRI to ``spin up'' and drive accretion \cite[for a more detailed analysis of the impact of magnetic field strength, see][]{Lund2023Beta}. Note also that the magnetic field topology also plays a role. More complex field loop structures are required to construct a MAD state and the MRI for a toroidal field loop is much slower than for poloidal.

The original motivation for this study was an attempt to understand the observation that GRB jet opening angles are narrower at higher redshifts.  If the {\em observed} jet opening angle was correlated with (or determined by) the size of the initial jet opening angle, the result would suggest, then, that GRB central engines are more rapidly rotating in the early universe relative to low-redshift GRBs. This is consistent with our current view of stellar evolution:  massive stars in the early universe are expected to have lower metallicity and therefore less radiation-driven winds. As a result, they experience less angular momentum loss, which may lead to a more rapidly rotating central engine upon collapse, and therefore a more collimated jet.  This explanation, of course, ignores the many intricacies of angular momentum transport/loss as the star collapses - it is still unclear to what extent we can connect the angular momentum of the progenitor at the end of its life with that of the black hole-disk system. 

This argument, however, does not account for the collimation by the stellar envelope and cocoon, which occurs at a distance much farther from the central engine than our work considers here. For example, \cite{Brom11, Brom14, Glob16} derive collimation conditions from the jet, cocoon and stellar envelope equilibrium pressures and \cite{Lloyd_Ronning_2020} showed, using these previous works, that this can explain the jet opening angle redshift correlation seen in the data, under certain conditions (i.e. an evolving initial mass function).  In other words, for jets that interact with a dense envelope (as expected in the case of a massive star collapse, but not necessarily in a neutron star merger), the observed jet opening angle is more likely a reflection of the interaction with this envelope rather than the initial jet opening angle set by the central engine.  Nonetheless, these results provide a promising avenue to further explore the imprint of the central engine on the jet physics. 
\\














\section{acknowledgments}

We thank the referee for valuable comments and suggestions that led to an improvement in this manuscript.  J.M.M. would like to thank J. Dolence, B. Ryan, P. Mullen for many helpful discussions, and the Institute for Nuclear Theory at the University of Washington for its kind hospitality and stimulating research environment.  N.M.L-R. would like to thank Roseanne Cheng, Gibwa Musoke, and Sera Markoff for elucidating conversations.  V.U.H would like to thank Kelly Holley-Bockelmann, Dina Stroud, Lauren Campbell, and the Fisk-Vanderbilt Master's to PhD Bridge Program for their mentorship over the years.
This work was supported through the Laboratory Directed Research and Development program under project number 20220564ECR at Los Alamos National Laboratory (LANL). 
LANL is operated by Triad National Security, LLC, for the National Nuclear Security Administration of U.S. Department of Energy (Contract No. 89233218CNA000001). 
This research used resources provided by the Los Alamos National Laboratory Institutional Computing Program.
This research was supported in part by the INT's U.S. Department of Energy grant No. DE-FG02- 00ER41132, and by the U.S. Department of Energy, Office of Science, Office of Workforce Development for Teachers and Scientists (WDTS) under the Science Undergraduate Laboratory Internships Program (SULI).
This work is approved for unlimited release with LA-UR-23-30405.

\bibliography{refs}{}

\begin{thebibliography}{}
\expandafter\ifx\csname natexlab\endcsname\relax\def\natexlab#1{#1}\fi
\providecommand{\url}[1]{\href{#1}{#1}}
\providecommand{\dodoi}[1]{doi:~\href{http://doi.org/#1}{\nolinkurl{#1}}}
\providecommand{\doeprint}[1]{\href{http://ascl.net/#1}{\nolinkurl{http://ascl.net/#1}}}
\providecommand{\doarXiv}[1]{\href{https://arxiv.org/abs/#1}{\nolinkurl{https://arxiv.org/abs/#1}}}

\bibitem[{{Balbus} \& {Hawley}(1991)}]{BalbusHawley91}
{Balbus}, S.~A., \& {Hawley}, J.~F. 1991, \apj, 376, 214, \dodoi{10.1086/170270}

\bibitem[{{Berger}(2014)}]{2014ARA&A..52...43B}
{Berger}, E. 2014, \araa, 52, 43, \dodoi{10.1146/annurev-astro-081913-035926}

\bibitem[{{Blandford} \& {Znajek}(1977{\natexlab{a}})}]{1977MNRAS.179..433B}
{Blandford}, R.~D., \& {Znajek}, R.~L. 1977{\natexlab{a}}, \mnras, 179, 433, \dodoi{10.1093/mnras/179.3.433}

\bibitem[{{Blandford} \& {Znajek}(1977{\natexlab{b}})}]{BZ77}
---. 1977{\natexlab{b}}, \mnras, 179, 433, \dodoi{10.1093/mnras/179.3.433}

\bibitem[{{Bromberg} {et~al.}(2014){Bromberg}, {Granot}, {Lyubarsky}, \& {Piran}}]{Brom14}
{Bromberg}, O., {Granot}, J., {Lyubarsky}, Y., \& {Piran}, T. 2014, \mnras, 443, 1532, \dodoi{10.1093/mnras/stu995}

\bibitem[{{Bromberg} {et~al.}(2011){Bromberg}, {Nakar}, {Piran}, \& {Sari}}]{Brom11}
{Bromberg}, O., {Nakar}, E., {Piran}, T., \& {Sari}, R. 2011, \apj, 740, 100, \dodoi{10.1088/0004-637X/740/2/100}

\bibitem[{{Christie} {et~al.}(2019){Christie}, {Lalakos}, {Tchekhovskoy}, {Fern{\'a}ndez}, {Foucart}, {Quataert}, \& {Kasen}}]{Chris19}
{Christie}, I.~M., {Lalakos}, A., {Tchekhovskoy}, A., {et~al.} 2019, \mnras, 490, 4811, \dodoi{10.1093/mnras/stz2552}

\bibitem[{{D'Avanzo}(2015)}]{DAvanz15}
{D'Avanzo}, P. 2015, Journal of High Energy Astrophysics, 7, 73, \dodoi{10.1016/j.jheap.2015.07.002}

\bibitem[{{Dav{\'e}}(2008)}]{Dave08}
{Dav{\'e}}, R. 2008, \mnras, 385, 147, \dodoi{10.1111/j.1365-2966.2008.12866.x}

\bibitem[{Dolence {et~al.}(2009)Dolence, Gammie, Mo{\'{s} }cibrodzka, \& Leung}]{Dolence_2009}
Dolence, J.~C., Gammie, C.~F., Mo{\'{s} }cibrodzka, M., \& Leung, P.~K. 2009, The Astrophysical Journal Supplement Series, 184, 387, \dodoi{10.1088/0067-0049/184/2/387}

\bibitem[{{Duncan} \& {Thompson}(1992)}]{DT92}
{Duncan}, R.~C., \& {Thompson}, C. 1992, \apjl, 392, L9, \dodoi{10.1086/186413}

\bibitem[{{Efron} \& {Petrosian}(1992)}]{EP92}
{Efron}, B., \& {Petrosian}, V. 1992, \apj, 399, 345, \dodoi{10.1086/171931}

\bibitem[{{Fishbone} \& {Moncrief}(1976)}]{1976ApJ...207..962F}
{Fishbone}, L.~G., \& {Moncrief}, V. 1976, \apj, 207, 962, \dodoi{10.1086/154565}

\bibitem[{{Frail} {et~al.}(2001){Frail}, {Kulkarni}, {Sari}, {Djorgovski}, {Bloom}, {Galama}, {Reichart}, {Berger}, {Harrison}, {Price}, {Yost}, {Diercks}, {Goodrich}, \& {Chaffee}}]{Frail01}
{Frail}, D.~A., {Kulkarni}, S.~R., {Sari}, R., {et~al.} 2001, \apjl, 562, L55, \dodoi{10.1086/338119}

\bibitem[{Gammie {et~al.}(2003)Gammie, McKinney, \& Toth}]{Gammie_2003}
Gammie, C.~F., McKinney, J.~C., \& Toth, G. 2003, The Astrophysical Journal, 589, 444–457, \dodoi{10.1086/374594}

\bibitem[{{Gehrels} {et~al.}(2009){Gehrels}, {Ramirez-Ruiz}, \& {Fox}}]{2009ARA&A..47..567G}
{Gehrels}, N., {Ramirez-Ruiz}, E., \& {Fox}, D.~B. 2009, \araa, 47, 567, \dodoi{10.1146/annurev.astro.46.060407.145147}

\bibitem[{{Globus} \& {Levinson}(2016)}]{Glob16}
{Globus}, N., \& {Levinson}, A. 2016, \mnras, 461, 2605, \dodoi{10.1093/mnras/stw1474}

\bibitem[{{Kerr}(1963)}]{1963PhRvL..11..237K}
{Kerr}, R.~P. 1963, \prl, 11, 237, \dodoi{10.1103/PhysRevLett.11.237}

\bibitem[{{Komissarov} {et~al.}(2009){Komissarov}, {Vlahakis}, {K{\"o}nigl}, \& {Barkov}}]{Kom09}
{Komissarov}, S.~S., {Vlahakis}, N., {K{\"o}nigl}, A., \& {Barkov}, M.~V. 2009, \mnras, 394, 1182, \dodoi{10.1111/j.1365-2966.2009.14410.x}

\bibitem[{{Laskar} {et~al.}(2018){Laskar}, {Berger}, {Chornock}, {Margutti}, {Fong}, \& {Zauderer}}]{Las18}
{Laskar}, T., {Berger}, E., {Chornock}, R., {et~al.} 2018, \apj, 858, 65, \dodoi{10.3847/1538-4357/aab8f5}

\bibitem[{{Laskar} {et~al.}(2014){Laskar}, {Berger}, {Tanvir}, {Zauderer}, {Margutti}, {Levan}, {Perley}, {Fong}, {Wiersema}, \& {Menten}}]{Las14}
{Laskar}, T., {Berger}, E., {Tanvir}, N., {et~al.} 2014, \apj, 781, 1, \dodoi{10.1088/0004-637X/781/1/1}

\bibitem[{{Lee} {et~al.}(2000){Lee}, {Brown}, \& {Wijers}}]{LWB00}
{Lee}, H.~K., {Brown}, G.~E., \& {Wijers}, R.~A.~M.~J. 2000, \apj, 536, 416, \dodoi{10.1086/308937}

\bibitem[{{Lee} \& {Ramirez-Ruiz}(2002)}]{LRR02}
{Lee}, W.~H., \& {Ramirez-Ruiz}, E. 2002, \apj, 577, 893, \dodoi{10.1086/342112}

\bibitem[{{Lee} \& {Ramirez-Ruiz}(2007)}]{2007NJPh....9...17L}
---. 2007, New Journal of Physics, 9, 17, \dodoi{10.1088/1367-2630/9/1/017}

\bibitem[{{Lei} {et~al.}(2017){Lei}, {Zhang}, {Wu}, \& {Liang}}]{Lei17}
{Lei}, W.-H., {Zhang}, B., {Wu}, X.-F., \& {Liang}, E.-W. 2017, \apj, 849, 47, \dodoi{10.3847/1538-4357/aa9074}

\bibitem[{{Levan} {et~al.}(2016){Levan}, {Crowther}, {de Grijs}, {Langer}, {Xu}, \& {Yoon}}]{Lev16}
{Levan}, A., {Crowther}, P., {de Grijs}, R., {et~al.} 2016, \ssr, 202, 33, \dodoi{10.1007/s11214-016-0312-x}

\bibitem[{{Liska} {et~al.}(2018){Liska}, {Hesp}, {Tchekhovskoy}, {Ingram}, {van der Klis}, \& {Markoff}}]{Lisk18}
{Liska}, M., {Hesp}, C., {Tchekhovskoy}, A., {et~al.} 2018, \mnras, 474, L81, \dodoi{10.1093/mnrasl/slx174}

\bibitem[{Lloyd-Ronning {et~al.}(2020)Lloyd-Ronning, Hurtado, Aykutalp, Johnson, \& Ceccobello}]{Lloyd_Ronning_2020}
Lloyd-Ronning, N., Hurtado, V.~U., Aykutalp, A., Johnson, J., \& Ceccobello, C. 2020, Monthly Notices of the Royal Astronomical Society, 494, 4371–4381, \dodoi{10.1093/mnras/staa1057}

\bibitem[{{Lloyd-Ronning} {et~al.}(2019{\natexlab{a}}){Lloyd-Ronning}, {Aykutalp}, \& {Johnson}}]{LR19}
{Lloyd-Ronning}, N.~M., {Aykutalp}, A., \& {Johnson}, J.~L. 2019{\natexlab{a}}, \mnras, 488, 5823, \dodoi{10.1093/mnras/stz2155}

\bibitem[{{Lloyd-Ronning} {et~al.}(2019{\natexlab{b}}){Lloyd-Ronning}, {Fryer}, {Miller}, {Prasad}, {Torres}, \& {Martin}}]{LR19b}
{Lloyd-Ronning}, N.~M., {Fryer}, C., {Miller}, J.~M., {et~al.} 2019{\natexlab{b}}, \mnras, 485, 203, \dodoi{10.1093/mnras/stz390}

\bibitem[{{Lloyd-Ronning} {et~al.}(2002){Lloyd-Ronning}, {Fryer}, \& {Ramirez-Ruiz}}]{LRFRR02}
{Lloyd-Ronning}, N.~M., {Fryer}, C.~L., \& {Ramirez-Ruiz}, E. 2002, \apj, 574, 554, \dodoi{10.1086/341059}

\bibitem[{L{\"u} {et~al.}(2012)L{\"u}, Zou, Lei, Zhang, Wu, Wang, Liang, \& L{\"u}}]{Lu12}
L{\"u}, J., Zou, Y.-C., Lei, W.-H., {et~al.} 2012, The Astrophysical Journal, 751, 49

\bibitem[{{Lund} {et~al.}(2023){Lund}, {McLaughlin}, {Miller}, \& {Mumpower}}]{Lund2023Beta}
{Lund}, K.~A., {McLaughlin}, G., {Miller}, J., \& {Mumpower}, M. 2023, arXiv e-prints, arXiv:2311.05796, \dodoi{10.48550/arXiv.2311.05796}

\bibitem[{{Lynden-Bell}(1971)}]{LB71}
{Lynden-Bell}, D. 1971, \mnras, 155, 95, \dodoi{10.1093/mnras/155.1.95}

\bibitem[{{Lyubarsky}(2009)}]{Lyub09}
{Lyubarsky}, Y. 2009, \apj, 698, 1570, \dodoi{10.1088/0004-637X/698/2/1570}

\bibitem[{{MacDonald} \& {Thorne}(1982)}]{MT82}
{MacDonald}, D., \& {Thorne}, K.~S. 1982, \mnras, 198, 345, \dodoi{10.1093/mnras/198.2.345}

\bibitem[{{MacFadyen} {et~al.}(2001){MacFadyen}, {Woosley}, \& {Heger}}]{2001ApJ...550..410M}
{MacFadyen}, A.~I., {Woosley}, S.~E., \& {Heger}, A. 2001, \apj, 550, 410, \dodoi{10.1086/319698}

\bibitem[{{McKinney} \& {Gammie}(2004)}]{McKinney2004}
{McKinney}, J.~C., \& {Gammie}, C.~F. 2004, \apj, 611, 977, \dodoi{10.1086/422244}

\bibitem[{{M{\'e}sz{\'a}ros}(2006)}]{Mesz06}
{M{\'e}sz{\'a}ros}, P. 2006, Reports on Progress in Physics, 69, 2259, \dodoi{10.1088/0034-4885/69/8/R01}

\bibitem[{{Mignone} \& {McKinney}(2007)}]{MignoneMckinney2007}
{Mignone}, A., \& {McKinney}, J.~C. 2007, \mnras, 378, 1118, \dodoi{10.1111/j.1365-2966.2007.11849.x}

\bibitem[{Miller {et~al.}(2019)Miller, Ryan, Dolence, Burrows, Fontes, Fryer, Korobkin, Lippuner, Mumpower, \& Wollaeger}]{Miller_2019}
Miller, J.~M., Ryan, B.~R., Dolence, J.~C., {et~al.} 2019, Physical Review D, 100, \dodoi{10.1103/physrevd.100.023008}

\bibitem[{{Narayan} {et~al.}(2022){Narayan}, {Chael}, {Chatterjee}, {Ricarte}, \& {Curd}}]{2022MNRAS.511.3795N}
{Narayan}, R., {Chael}, A., {Chatterjee}, K., {Ricarte}, A., \& {Curd}, B. 2022, \mnras, 511, 3795, \dodoi{10.1093/mnras/stac285}

\bibitem[{{Pais} {et~al.}(2023){Pais}, {Piran}, {Lyubarsky}, {Kiuchi}, \& {Shibata}}]{Pais2023}
{Pais}, M., {Piran}, T., {Lyubarsky}, Y., {Kiuchi}, K., \& {Shibata}, M. 2023, \apjl, 946, L9, \dodoi{10.3847/2041-8213/acc2c5}

\bibitem[{{Piran}(2004)}]{2004RvMP...76.1143P}
{Piran}, T. 2004, Reviews of Modern Physics, 76, 1143, \dodoi{10.1103/RevModPhys.76.1143}

\bibitem[{{Porth} {et~al.}(2019){Porth}, {Chatterjee}, {Narayan}, {Gammie}, {Mizuno}, {Anninos}, {Baker}, {Bugli}, {Chan}, {Davelaar}, {Del Zanna}, {Etienne}, {Fragile}, {Kelly}, {Liska}, {Markoff}, {McKinney}, {Mishra}, {Noble}, {Olivares}, {Prather}, {Rezzolla}, {Ryan}, {Stone}, {Tomei}, {White}, {Younsi}, {Akiyama}, {Alberdi}, {Alef}, {Asada}, {Azulay}, {Baczko}, {Ball}, {Balokovi{\'c}}, {Barrett}, {Bintley}, {Blackburn}, {Boland}, {Bouman}, {Bower}, {Bremer}, {Brinkerink}, {Brissenden}, {Britzen}, {Broderick}, {Broguiere}, {Bronzwaer}, {Byun}, {Carlstrom}, {Chael}, {Chatterjee}, {Chen}, {Chen}, {Cho}, {Christian}, {Conway}, {Cordes}, {Geoffrey}, {Crew}, {Cui}, {De Laurentis}, {Deane}, {Dempsey}, {Desvignes}, {Doeleman}, {Eatough}, {Falcke}, {Fish}, {Fomalont}, {Fraga-Encinas}, {Freeman}, {Friberg}, {Fromm}, {G{\'o}mez}, {Galison}, {Garc{\'\i}a}, {Gentaz}, {Georgiev}, {Goddi}, {Gold}, {Gu}, {Gurwell}, {Hada}, {Hecht}, {Hesper}, {Ho}, {Ho}, {Honma}, {Huang}, {Huang}, {Hughes}, {Ikeda}, {Inoue}, {Issaoun},
  {James}, {Jannuzi}, {Janssen}, {Jeter}, {Jiang}, {Johnson}, {Jorstad}, {Jung}, {Karami}, {Karuppusamy}, {Kawashima}, {Keating}, {Kettenis}, {Kim}, {Kim}, {Kim}, {Kino}, {Koay}, {Patrick}, {Koch}, {Koyama}, {Kramer}, {Kramer}, {Krichbaum}, {Kuo}, {Lauer}, {Lee}, {Li}, {Li}, {Lindqvist}, {Liu}, {Liuzzo}, {Lo}, {Lobanov}, {Loinard}, {Lonsdale}, {Lu}, {MacDonald}, {Mao}, {Marrone}, {Marscher}, {Mart{\'\i}-Vidal}, {Matsushita}, {Matthews}, {Medeiros}, {Menten}, {Mizuno}, {Moran}, {Moriyama}, {Moscibrodzka}, {M{\"u}ller}, {Nagai}, {Nagar}, {Nakamura}, {Narayanan}, {Natarajan}, {Neri}, {Ni}, {Noutsos}, {Okino}, {Oyama}, {{\"O}zel}, {Palumbo}, {Patel}, {Pen}, {Pesce}, {Pi{\'e}tu}, {Plambeck}, {PopStefanija}, {Preciado-L{\'o}pez}, {Psaltis}, {Pu}, {Ramakrishnan}, {Rao}, {Rawlings}, {Raymond}, {Ripperda}, {Roelofs}, {Rogers}, {Ros}, {Rose}, {Roshanineshat}, {Rottmann}, {Roy}, {Ruszczyk}, {Rygl}, {S{\'a}nchez}, {S{\'a}nchez-Arguelles}, {Sasada}, {Savolainen}, {Schloerb}, {Schuster}, {Shao}, {Shen}, {Small}, {Sohn},
  {SooHoo}, {Tazaki}, {Tiede}, {Tilanus}, {Titus}, {Toma}, {Torne}, {Trent}, {Trippe}, {Tsuda}, {van Bemmel}, {van Langevelde}, {van Rossum}, {Wagner}, {Wardle}, {Weintroub}, {Wex}, {Wharton}, {Wielgus}, {Wong}, {Wu}, {Young}, {Young}, {Yuan}, {Yuan}, {Zensus}, {Zhao}, {Zhao}, {Zhu}, \& {Event Horizon Telescope Collaboration}}]{EHTComparison2019}
{Porth}, O., {Chatterjee}, K., {Narayan}, R., {et~al.} 2019, \apjs, 243, 26, \dodoi{10.3847/1538-4365/ab29fd}

\bibitem[{{Ressler} {et~al.}(2015){Ressler}, {Tchekhovskoy}, {Quataert}, {Chandra}, \& {Gammie}}]{Ressler2015ebhlight}
{Ressler}, S.~M., {Tchekhovskoy}, A., {Quataert}, E., {Chandra}, M., \& {Gammie}, C.~F. 2015, \mnras, 454, 1848, \dodoi{10.1093/mnras/stv2084}

\bibitem[{{Rohoza} {et~al.}(2023){Rohoza}, {Lalakos}, {Paik}, {Chatterjee}, {Liska}, {Tchekhovskoy}, \& {Gottlieb}}]{Roh23}
{Rohoza}, V., {Lalakos}, A., {Paik}, M., {et~al.} 2023, arXiv e-prints, arXiv:2311.00018, \dodoi{10.48550/arXiv.2311.00018}

\bibitem[{Ryan {et~al.}(2015)Ryan, Dolence, \& Gammie}]{ryan2015bhlight}
Ryan, B.~R., Dolence, J.~C., \& Gammie, C.~F. 2015, bhlight: General Relativistic Radiation Magnetohydrodynamics with Monte Carlo Transport.
\newblock \doarXiv{1505.05119}

\bibitem[{{Sano} {et~al.}(2004){Sano}, {Inutsuka}, {Turner}, \& {Stone}}]{Sano2004}
{Sano}, T., {Inutsuka}, S.-i., {Turner}, N.~J., \& {Stone}, J.~M. 2004, \apj, 605, 321, \dodoi{10.1086/382184}

\bibitem[{{Shakura} \& {Sunyaev}(1973)}]{ShakuraSunyaev1973}
{Shakura}, N.~I., \& {Sunyaev}, R.~A. 1973, \aap, 24, 337

\bibitem[{{Shiokawa} {et~al.}(2012){Shiokawa}, {Dolence}, {Gammie}, \& {Noble}}]{Shiokawa2011}
{Shiokawa}, H., {Dolence}, J.~C., {Gammie}, C.~F., \& {Noble}, S.~C. 2012, \apj, 744, 187, \dodoi{10.1088/0004-637X/744/2/187}

\bibitem[{{Tchekhovskoy} \& {McKinney}(2012)}]{tm12}
{Tchekhovskoy}, A., \& {McKinney}, J.~C. 2012, \mnras, 423, L55, \dodoi{10.1111/j.1745-3933.2012.01256.x}

\bibitem[{{Tchekhovskoy} {et~al.}(2007){Tchekhovskoy}, {McKinney}, \& {Narayan}}]{Tchekhovskoy2007}
{Tchekhovskoy}, A., {McKinney}, J.~C., \& {Narayan}, R. 2007, \mnras, 379, 469, \dodoi{10.1111/j.1365-2966.2007.11876.x}

\bibitem[{{Tchekhovskoy} {et~al.}(2012){Tchekhovskoy}, {McKinney}, \& {Narayan}}]{tmn12}
{Tchekhovskoy}, A., {McKinney}, J.~C., \& {Narayan}, R. 2012, in Journal of Physics Conference Series, Vol. 372, Journal of Physics Conference Series, 012040, \dodoi{10.1088/1742-6596/372/1/012040}

\bibitem[{{Thompson}(1994)}]{Thomp94}
{Thompson}, C. 1994, \mnras, 270, 480, \dodoi{10.1093/mnras/270.3.480}

\bibitem[{{Usov}(1992)}]{Usov92}
{Usov}, V.~V. 1992, \nat, 357, 472, \dodoi{10.1038/357472a0}

\bibitem[{Velikhov(1959)}]{VelikovMRI}
Velikhov, E. 1959, Zhur. Eksptl'. i Teoret. Fiz., 36

\bibitem[{{Woosley} \& {Bloom}(2006)}]{2006ARA&A..44..507W}
{Woosley}, S.~E., \& {Bloom}, J.~S. 2006, \araa, 44, 507, \dodoi{10.1146/annurev.astro.43.072103.150558}

\bibitem[{{Woosley} \& {Heger}(2006)}]{2006ApJ...637..914W}
{Woosley}, S.~E., \& {Heger}, A. 2006, \apj, 637, 914, \dodoi{10.1086/498500}

\bibitem[{{Woosley} \& {Heger}(2012)}]{2012ApJ...752...32W}
---. 2012, \apj, 752, 32, \dodoi{10.1088/0004-637X/752/1/32}

\bibitem[{{Yonetoku} {et~al.}(2005){Yonetoku}, {Yamazaki}, {Nakamura}, \& {Murakami}}]{Yon05}
{Yonetoku}, D., {Yamazaki}, R., {Nakamura}, T., \& {Murakami}, T. 2005, \mnras, 362, 1114, \dodoi{10.1111/j.1365-2966.2005.09398.x}

\bibitem[{{Zhang} \& {M{\'e}sz{\'a}ros}(2001)}]{ZM01}
{Zhang}, B., \& {M{\'e}sz{\'a}ros}, P. 2001, \apjl, 552, L35, \dodoi{10.1086/320255}

\bibitem[{{Zhang} \& {M{\'e}sz{\'a}ros}(2004)}]{ZM04}
---. 2004, International Journal of Modern Physics A, 19, 2385, \dodoi{10.1142/S0217751X0401746X}

\end{thebibliography}
\bibliographystyle{aasjournal}



\end{document}